\documentclass[english,a4paper]{jpconf}
\usepackage[T1]{fontenc}
\usepackage[latin9]{inputenc}
\usepackage{babel}
\usepackage{units}
\usepackage[unicode=true,pdfusetitle,
 bookmarks=true,bookmarksnumbered=false,bookmarksopen=false,
 breaklinks=false,pdfborder={0 0 1},backref=false,colorlinks=false]
 {hyperref}
\usepackage{breakurl}

\makeatletter
\usepackage{enumitem}		% customizable list environments
\newlength{\lyxlabelwidth}      % auxiliary length 
\newenvironment{elabeling}[2][]%
{\settowidth{\lyxlabelwidth}{#2}
\begin{description}[font=\normalfont,style=sameline,
leftmargin=\lyxlabelwidth,#1]}
{\end{description}}

\usepackage{cite}

\newcommand{\ttbar}{\ensuremath{\rm{t}\overline{\rm{t}}}~} % t-tbar

\renewenvironment{elabeling}[2][]%
{\settowidth{\lyxlabelwidth}{#2}
\footnotesize \vspace{-0.75em}
\begin{description}[itemsep=-0.25em, font=\normalfont, style=sameline, leftmargin=\lyxlabelwidth,#1]}
{\end{description} \normalsize \vspace{-0.75em}}

\makeatother

\begin{document}
\noindent \title{Model uncertainties in top-quark physics}
%\vspace{-2.5cm}
\author{Markus Seidel}
\address{Universit\"at Hamburg, on behalf of the ATLAS and CMS collaborations}
\ead{markus.seidel@uni-hamburg.de}
\begin{abstract}
\noindent The ATLAS and CMS collaborations at the Large Hadron Collider
(LHC) are studying the top quark in pp collisions at $\sqrt{s}=7\mbox{ and }8\mbox{ TeV}$.
Due to the large integrated luminosity, precision measurements of
production cross-sections and properties are often limited by systematic
uncertainties. An overview of the modeling uncertainties for simulated
events is given in this report.
\end{abstract}

\section{Introduction}

The successful Run 1 of the LHC allowed the ATLAS~\cite{Aad:2008zzm}
and CMS~\cite{Chatrchyan:2008aa} collaborations to acquire pp collision
datasets at $\sqrt{s}=7\mbox{ and }8\mbox{ TeV}$ corresponding to
$25\mbox{ fb}^{-1}$. This lead to a reduction of the statistical
uncertainties in top-quark measurements, especially in combined results~\cite{ATLAS:2014wva,CMS-PAS-TOP-12-002}.
Among the systematic uncertainties, the ones on the modeling of top-quark
pair and single-top events contribute significantly to the precision
limit. Simulated events are predictions of the Standard Model of particle
physics (or a theory beyond that) using Monte-Carlo (MC) methods,
and consist out of matrix-element calculation, parton shower and hadronization
stage, implemented in various publicly available codes~\cite{Buckley:2011ms,Alwall:2014hca,Alioli:2010xd,Frixione:2002ik,Mangano:2002ea,Gleisberg:2008ta,Boos:2004kh,Kersevan:2004yg,Corcella:2000bw,Bahr:2008pv,Sjostrand:2006za,Sjostrand:2014zea,Lange:2001uf}.
These models are tuned to data from both current and preceeding experiments.
The resulting stable particles (photons, leptons, and hadrons) can
be passed to a detector simulation in order to compare to data. This
simulation chain allows for the determination of unobservable parameters
(as the top-quark mass) or for the estimation of signal selection
efficiencies, needed for cross-section measurements. This report discusses
the most relevant uncertainties from perturbative QCD in Sec. \ref{sec:Perturbative-QCD-uncertainties}
and those from soft QCD in Sec.\ref{sec:Soft-QCD-modeling}, summarizing
the prescriptions employed in ATLAS and CMS.

\section{Perturbative QCD uncertainties\label{sec:Perturbative-QCD-uncertainties}}

\subsection{Parton density functions}

The uncertainty on the parton density function of the proton is evaluated
using the PDF4LHC prescription, creating an envelope of three PDF
sets and their respective uncertainties~\cite{Botje:2011sn}. Including
a variation of the strong coupling $\alpha_{s}$, a set of 147 PDFs
is evaluated, becoming a bottle-neck for CPU-intensive analyses like
matrix-element methods. Therefore, for insensitive analyses often
only the variations of one PDF set are used and compared to the central
values of the other PDFs. A prescription for covering all PDF sets
and their uncertainties with a reduced number of variations would
be desirable from experimental point of view, possibly in the framework
of a meta-analysis~\cite{Gao:2013bia}.
\begin{elabeling}{00.00.0000}
\item [{ATLAS}] PDF4LHC prescription, or uncertainties of default PDF +
other central PDFs
\item [{CMS}] PDF4LHC prescription, or uncertainties of default PDF
\end{elabeling}

\subsection{$t\overline{t}$ MC generator}

The guidelines of the Top LHC working group~\cite{TopLHCWG} recommend
the comparison of central predictions from different MC generators,
using at least one multi-leg and one NLO generator setup. Additional
uncertainties are estimated by parameter variations inside a generator
framework to disentangle different effects.
\begin{elabeling}{00.00.0000}
\item [{ATLAS}] Powheg+Pythia6 vs. MC@NLO+Herwig6 (vs. Alpgen+Herwig6)
\item [{CMS}] MadGraph+Pythia6 vs. Powheg+Pythia6
\end{elabeling}

\subsection{Single-top MC generator}

Simulating single-top t-channel events is possible in either the 5-
or the 4-flavour scheme (FS). The 5FS is based on massless b quarks
in the proton PDF, reducing the complexity of the LO calculation.
The 4FS contains the $g\rightarrow b\bar{b}$ splitting, yielding
a ME description of the additional b quark in the event. Matched schemes,
as implemented in AcerMC and CompHep, combine the 5FS and 4FS LO diagrams,
and NLO generators provide matching of diagrams in either 5FS or 4FS
(Powheg, aMCatNLO).
\begin{elabeling}{00.00.0000}
\item [{ATLAS}] AcerMC+Pythia6 (matched LO) vs. aMC@NLO+Herwig6 (4FS NLO)
\item [{CMS}] Powheg vs. aMC@NLO (4FS NLO)
\end{elabeling}
At NLO, single-top production in the tW channel overlaps with \ttbar
production. The diagram-removal scheme removes the double-resonant
diagrams from the signal definition, while the diagram-subtraction
scheme implements a subtraction term cancelling the \ttbar contribution
locally~\cite{Frixione:2008yi,Tait:1999cf}. There is on-going work
on a consistent treatment as $WWb\bar{b}$ final state, including
non-, single-, and double-resonant contributions, and quantum interference
effects~\cite{Frederix:2013gra,Cascioli:2013wga}.
\begin{elabeling}{00.00.0000}
\item [{ATLAS \& CMS}] Diagram removal vs. diagram subtraction
\end{elabeling}

\subsection{Radiation}

The amount of QCD radiation in an event affects top-quark reconstruction
and selection efficiencies. Theory-inspired scale variations by factors
of $\nicefrac{1}{2}$ and $2$ are found to be generous envelopes
of jet multiplicity and gap fraction measurements in \ttbar events
performed by ATLAS~\cite{Aad:2014iaa,ATL-PHYS-PUB-2014-005} and
CMS~\cite{Chatrchyan:2014gma,CMS-PAS-TOP-12-041}. Final-state radiation
inside resonance decays is tightly constrained by measurements of
event shapes at LEP~\cite{Heister:2003aj,Abdallah:2003xz,Achard:2004sv,Abbiendi:2004qz},
and is validated by the ATLAS measurement of jet shapes in \ttbar
events~\cite{Aad:2013fba}.
\begin{elabeling}{00.00.0000}
\item [{ATLAS}] Variation of renormalization scales in Alpgen+Pythia6
\item [{CMS}] Variation of renormalization and factorization scales and
ME-PS matching threshold in MadGraph+Pythia6
\end{elabeling}

\subsection{Momentum reshuffling and top-quark transverse momentum}

Differential \ttbar cross-section measurements indicate a softer
transverse momentum ($p_{{\rm T}}$) distribution than most predictions~\cite{Aad:2014zka,Chatrchyan:2012saa,CMS-PAS-TOP-12-027,CMS-PAS-TOP-12-028}.
However, Powheg+Herwig6 simulation shows good agreement in top-quark
$p_{{\rm T}}$ due to its momentum reshuffling scheme~\cite{Gorner:2014lpa,Nason}.
In the \ttbar NLO matrix-element used by Powheg, the real-radiation
parton is generated with zero mass. Interfacing to a parton shower
requires to assign virtuality to the additional parton, accomplished
by rescaling the momenta of the top-quarks and the parton by a common
factor. Other schemes are possible, like rescaling the momenta of
the parton and the \ttbar system, yielding a harder top-quark $p_{{\rm T}}$,
and are implemented in Herwig++. Pythia8 allows to compare a dipole-recoil
and a global recoil scheme.
\begin{elabeling}{00.00.0000}
\item [{CMS}] Reweight top-quark $p_{{\rm T}}$ to CMS measurement, assign
full difference as uncertainty
\end{elabeling}

\section{Soft QCD modeling\label{sec:Soft-QCD-modeling}}

\subsection{Hadronization model}

After parton shower evolution down to a cutoff scale in the order
of $1\mbox{ GeV}$, hadrons are formed via Lund string~\cite{Andersson:1983ia}
or cluster fagmentation~\cite{Webber:1983if,Winter:2003tt} models.
Similar tunings to $e^{+}e^{-}$ data may lead to different predictions
in top-quark events~\cite{ATL-PHYS-PUB-2014-008}, where the detector
response depends on momenta and types of stable particles. The comparison
of the reference implementations in Pythia and Herwig is non-trivial
due to further differences in parton shower, matching and underlying
event, that may add up or compensate each other. A comparison of string
and cluster fragmentation in Sherpa shows good agreement of parton$\to$particle
jet response and reconstructed top-quark and W-boson masses at particle
level~\cite{Seidel}.
\begin{elabeling}{00.00.0000}
\item [{ATLAS}] Powheg+Pythia6 vs. Powheg+Herwig6 in top-quark events,
and Pythia6 vs. Herwig++ in (b) jet energy scale~\cite{Aad:2014bia}
\item [{CMS}] Pythia6 vs. Herwig++ as fully flavor-dependent jet energy
scale uncertainty~\cite{Chatrchyan:2011ds,CMS-DP-2013-033}\\
Cross-checked by measurement of the b-jet energy scale in Z+b events~\cite{CMS-PAS-JME-13-001},
and Powheg+Pythia6 vs. MC@NLO/Powheg+Herwig6 comparison in top-quark
events
\end{elabeling}

\subsection{b fragmentation}

The fragmentation of bottom quarks influences b-tag efficiencies and
b-jet energy response. The parameters of the fragmentation function
are tuned to measurements of $x_{B}$ in $e^{+}e^{-}$ collisions~\cite{Heister:2001jg,Barker:994376,Abe:2002iq},
where $x_{B}=E_{B}/E_{beam}$ and $B$ denotes the weakly decaying
B hadron. Different tunings of the Bowler-Lund fragmentation function
in Pythia are used to evaluate the uncertainy. The identification
of charmed mesons inside b jets in \ttbar events and the measurement
of their momentum fractions establish a first step to validation of
b fragmentation at the LHC~\cite{CMS-PAS-TOP-13-007}.
\begin{elabeling}{00.00.0000}
\item [{ATLAS}] Pythia6 AMBT1 vs. P11 vs. ``Bowler modified'' tunes in
b jet energy scale
\item [{CMS}] Pythia6 Z2{*} vs. Z2{*}rbLEP tunes in top-quark events
\end{elabeling}

\subsection{B-hadron decays}

Measurements are affected by the uncertainty on B-hadron lifetime
either directly by a lifetime-based observable or indirectly by b-tag
efficiencies. As b-tag efficiencies are calibrated using data, no
additional uncertainty is quoted for most measurements.

The branching ratio of B hadrons decaying via $B\to\ell\nu X$ has
a direct impact on the fraction of undetected neutrinos inside b jets
and thus on the b-jet response.
\begin{elabeling}{00.00.0000}
\item [{CMS}] Envelope of PDG values for $B^{+}$ and $B^{0}$ semi-leptonic
branching ratios~\cite{Agashe:2014kda}
\end{elabeling}
Pythia6 and Herwig6 contain reduced decay tables, and will be replaced
by their successors Pythia8 and Herwig++ for the LHC Run 2. These
contain improved decay tables, similar to those included in Sherpa
and EvtGen.

\subsection{Underlying event}

A hard scattering process is accompanied by additional parton interactions
of the protons, called the underlying event. Its description by event
generators is tuned to particle spectra in minimum bias events. The
origin of charged particles from the underlying event cannot be distinguished
from the primary interaction vertex detected in experiment, so that
their energy is added to clustered jets. Pile-up mitigation techniques
based on particle densities may still compensate for variations of
the underlying event activity and shape.
\begin{elabeling}{00.00.0000}
\item [{ATLAS}] Pythia6 P11 vs. P11mpiHi, or P12 vs. P12mpiHi tunes
\item [{CMS}] Pythia6 P11 vs. P11mpiHi and P11TeV tunes
\end{elabeling}

\subsection{Color reconnection}

Color reconnection models allow non-perturbative changes in the color
configuration of the event, typically reducing the total potential
energy between QCD color charges~\cite{Sandhoff:2005jh}. This mechanism
improves the description of of charged particle $\left\langle p_{T}\right\rangle $
vs. $N_{ch}$ in minimum-bias events, although the current models
are not able to give a consistent tune to data in all $p_{T}$ ranges.
A new set of color reconnection models was recently implemented in
Pythia8~\cite{Argyropoulos:2014zoa}.
\begin{elabeling}{00.00.0000}
\item [{ATLAS}] Pythia6 P11 vs. P11noCR, or P12 vs. P12loCR tunes
\item [{CMS}] Pythia6 P11 vs. P11noCR tunes
\end{elabeling}

\section{Summary and outlook}

Simulated events are an important ingredient to LHC data analysis.
The models need to be tuned to reference data in order to get accurate
predictions for phase-space regions opening up at the LHC. A number
of careful parameter variations in the simulation programs is performed
to get a reliable estimate of the modeling uncertainties. There is
on-going and successful work in harmonizing the different prescriptions
used by ATLAS and CMS within the Top LHC working group, mainly driven
by combination efforts.

For the LHC Run 2, new NLO+multileg generators are expected reduce
the uncertainties on perturbative QCD~\cite{Alwall:2014hca,Hoeche:2014qda}.
At the same time, the inclusion of weights for generator variations
in both matrix element and parton shower has potential for efficient
and precise estimation of the uncertainties, without dilution by limited
statistical precision~\cite{Frederix:2011ss,Hartgring:2013jma}.

Complementary analysis strategies are being followed to preserve ATLAS
and CMS data for all practical purposes. New cross-section measurements
quote results also in fiducial phase spaces that are closer to the
detector acceptance~\cite{ATLAS-CONF-2014-007}. The extrapolation
from fiducial to inclusive cross sections will be possible using any
improved calculation in the future, benefitting from reduced uncertainties.
Definition of the top quark at particle level will reduce the modeling
uncertainties on differential measurements~\cite{ATLAS-CONF-2014-059,CMS-PAS-TOP-12-028}
and allow for generator comparison and tuning in the Rivet+Professor
framework~\cite{Buckley:2010ar,Buckley:2009bj}.

\section*{\noindent References}

\noindent \bibliographystyle{iopart-num}
\bibliography{top_modeling_mseidel}

\end{document}